\documentclass[aps,pra,preprint,groupedaddress,showpacs]{revtex4}
\usepackage{graphics}
\usepackage{bm}
\begin{document}

\title{Paraxial propagation in amorphous optical media with screw dislocation}                                                                    

\author{Leila Mashhadi, Mohammad Mehrafarin}
\email{mehrafar@aut.ac.ir}
\affiliation{Physics Department, Amirkabir University of Technology, Tehran 15914, Iran}                                                                              
\begin{abstract}
We study paraxial beam propagation parallel to the screw axis of a dislocated amorphous medium that is optically weakly inhomogeneous and isotropic. The effect of the screw dislocation on the beam's orbital angular momentum is shown to change the optical vortex strength, rendering vortex annihilation or generation possible.  Furthermore, the dislocation is shown to induce a weak \textit{biaxial} anisotropy in the medium due to the elasto-optic effect, which changes the beam's spin angular momentum as well as causing precession of the polarization. We derive the equations of motion of the beam and demonstrate the optical Hall effect in the dislocated medium. Its application with regard to determining the Burgers vector as well as the elasto-optic coefficients of the medium is explained.
\end{abstract}
                                              
\maketitle
\section{Introduction}

Paraxial beams with non-zero orbital angular momentum, which are exact eigenmodes of the paraxial wave equation \cite{Saleh}, have been of considerable interest \cite{Allen,Allen2,Maier}. In addition to spin angular momentum (polarization), paraxial beams carry an intrinsic orbital angular momentum that depends on their spatial structure. While the transfer of spin angular momentum to matter has long been reported for optically anisotropic media \cite{Beth}, the transfer of orbital angular momentum has newly triggered a wide range of applications. A notable example is particle trapping \cite{Simpson}, whereby particles are set into rotation by coupling with the beam's orbital angular momentum and are thus trapped. Such a coupling of the orbital angular momentum with matter is possible in optically inhomogeneous isotropic media \cite{Beijersbergen}. Thus, a simultaneous independent coupling of both spin and orbital angular momentum with matter is to be expected in a medium which is both optically inhomogeneous and anisotropic \cite{Piccirillo,Marrucci}. 

In the past decades, the geometric Berry phase \cite{Berry} acquired by an optical beam (as the Pancharatnam phase \cite{Pancharatnam} or a spin redirection phase \cite{Tomita,Chiao}) has attracted extensive attention. In particular, Bliokh et. al. \cite{Bliokh,Bliokh1,Bliokh2} developed a modified geometrical optics approximation for optically weakly inhomogeneous isotropic media, which contained the Rytov-Vladimirskii rotation law \cite{Rytov,Vladimirskii} and the optical Magnus effect \cite{Zeldovich,Zeldovich2} (also called the optical Hall effect) as manifestations of Berry effects. Such geometric Berry effects have been also derived via the semiclassical wave packet approximation \cite{Onoda,Onoda2,Sawada}, which was originally developed for the study of electronic spin transport in solids \cite{Chang,Sundaram,Culcer}. 

By studying the propagation of paraxial beams in optically weakly inhomogeneous isotropic media, Bliokh \cite{Bliokh4} demonstrated that the optical Hall effect becomes more pronounced because of the contribution of the orbital angular momentum Hall effect due to the orbit-orbit interaction. Here, we generalize his results for screw dislocated (amorphous) optical media. Mechanical deformation can result in dislocations that change the optical properties of a medium due to the elasto-optic effect. Such dislocations can be induced by the focused beam itself (see e.g. \cite{Kanehira,Brandstetter}). The study of beam transport provides an indirect way to determine the dislocations as well as the elasto-optical properties of the medium. In the present work, we study paraxial beam propagation parallel to the screw axis of a dislocated amorphous medium that is optically weakly inhomogeneous and isotropic. The paraxial beam contains an optical dislocation, namely the optical vortex along its axis, which is a topological object on the wavefront the beam \cite{Berry2}. Regarding the medium as a continuum, we use the differential geometric theory of defects which is isomorphic to three dimensional gravity \cite{Kleinert,Katanaev}. In this geometric approach, the effect of defects on the three dimensional geometry of the medium is incorporated into a metric. The effect of the screw dislocation on the beam's orbital angular momentum is, thus, shown to change the optical vortex strength, rendering vortex annihilation (for left-handed helical modes) or generation of non-integral vortices \cite{Gotte} possible. This is useful in generating fast switchable helical modes for optical information encoding \cite{Marrucci,Marrucci2}. Furthermore, the dislocation is shown to induce a weak \textit{biaxial} anisotropy in the medium due to the elasto-optic effect. This anisotropy, which is felt more strongly by beams that propagate close to the dislocation line, changes the beam's spin angular momentum as well as causing precession of the polarization. Finally, we derive the equations of motion of the beam and demonstrate how beams with different values of spin and/or angular momentum split in the dislocated medium (the optical Hall effect). Because the beam's singular vortex core can be observed with great accuracy, measurement of the splittings is very reliable. Measuring beam splittings for opposite values of the polarization and orbital angular momentum provides an indirect method for determining the Burgers vector as well as the elasto-optic coefficients of the medium. Determination of the former is demonstrated in a simple example.

\section{Effect of screw dislocation on the beam}

Dislocation is a defect caused by the action of internal stress, which changes the physical properties of the medium, in particular, its optical properties. The displacement vector field, $\textbf{u}$, associated with a screw dislocation line oriented along the $z$-axis of the cylindrical coordinates $(\rho,\varphi,z)$ is $\textbf{u}=(0,0,\beta\varphi)$, where $\beta$ is the magnitude of Burgers vector divided by $2\pi$. The corresponding strain tensor field $S_{ij}=\frac{1}{2}(\partial_i u_j+\partial_j u_i)$ is, thus, given by $S_{\varphi z}=S_{z\varphi}=\beta/2\rho$, other components being zero. Regarding the medium as a continuum, we can use the differential geometric theory of defects which is isomorphic to three dimensional gravity \cite{Kleinert,Katanaev}. In this geometric approach, the effect of defects on the three dimensional geometry of the medium is incorporated into a metric. In particular, a screw dislocation, which corresponds to a singular torsion along the defect line, is described by the following metric \cite{Tod}
\begin{equation}
ds^2=d\rho^2+\rho^2 d\varphi^2+(dz+\beta d\varphi)^2. \label{metric}
\end{equation} 
This metric carries torsion but no curvature and describes a screw dislocated medium. 

We consider a monochromatic paraxial beam with definite values of the spin and orbital angular momentum, propagating in a weakly inhomogeneous isotropic medium where the refractive index $n(\textbf{x})$ varies adiabatically with position ($\nabla n \rightarrow 0$). With the screw axis $z$ as the paraxial direction, the beam's electric field can be represented as in a homogeneous medium according to 
\begin{equation}
\textbf{E}_{\sigma lm}(\rho,\varphi,z)=\textbf{e}_\sigma E_{lm}(\rho,\varphi)e^{ik z}. \label{soln}
\end{equation}
Here $k=k_0n$ is the wave number, the adiabatic variation of which has been neglected ($k_0$ being the wave number in vacuum), 
$$\textbf{e}_\sigma =\frac{1}{\surd 2}(\textbf{e}_\rho  -i\sigma \textbf{e}_\varphi)$$
with $\sigma=\pm1$ ($\textbf{e}_\rho , \textbf{e}_\varphi$ being the cylindrical unit vectors) are the orthonormal polarization vectors corresponding, respectively, to right/left circular polarization and
$$
E_{lm}(\rho,\varphi)=R_{l|m|}(\rho) e^{im\varphi} 
$$
$R_{l|m|}$; with $l=0,1,2,\ldots$ and $m=-l,-l+1,\ldots, l$; being the radial solution of the Helmholtz equation ($\sigma$  and $m$ represent the helicity and the $z$-component of the orbital angular momentum of the photon, respectively ($\hbar=1$)). In the presence of screw dislocation, the Laplacian operator $\nabla^2$ appearing in the Helmholtz equation is to be replaced by
$$
\partial_z^2+\frac{1}{\rho}\partial_\rho(\rho\partial_\rho)+\frac{1}{\rho^2}(\partial_\varphi-\beta\partial_z)^2
$$
which is the Laplacian associated with the metric (\ref{metric}) of the dislocated medium. The solution (\ref{soln}), therefore, acquires an additional phase factor due to the coupling of torsion with angular momentum, according to 
\begin{equation}
E_{lm}(\rho,\varphi)= R_{l|m|}(\rho) e^{i(m+\beta k)\varphi}.\label{soln2}
\end{equation}
Furthermore, the relative permittivity tensor $n^2 \delta_{ij}$ acquires an anisotropic part $\Delta_{ij}$ due to the strain field of the dislocation, where (see e.g. \cite{Liu}) 
$$
\Delta_{ij}=-n^4 p_{ijkl}S_{kl}
$$ 
$p_{ijkl}$ being the elasto-optic coefficients of the medium. For amorphous media, where only two independent elasto-optic coefficients (customarily denoted by $p_{11}$ and $p_{12}$) exist, the screw dislocation yields
$$
\Delta_{\varphi z}=\Delta_{z\varphi}=-\frac{\beta (p_{11}-p_{12})}{\rho}n^4
$$
other components being zero. The principle refractive indices are, therefore,
$$
n_\rho= n,\ \ n_\varphi=n+\frac{\beta (p_{11}-p_{12})}{2\rho}n^3,\ \ n_z=n-\frac{\beta (p_{11}-p_{12})}{2\rho}n^3
$$
to first order in the perturbation, which is generally sufficiently small. Hence, the elasto-optic perturbation induces a weak \textit{biaxial} anisotropy in the medium, which is felt more strongly by beams that propagate close to the dislocation line (small $\rho$). Propagation of the beam through the dislocated medium, thus, introduces the phases $k_0 n_\rho z$ and $k_0 n_\varphi z$ in the linear polarization states represented by $\textbf{e}_\rho$ and $\textbf{e}_\varphi$, respectively. Hence, in (\ref{soln}), the polarization vector should be rewritten as
$$
\textbf{e}_\sigma =\frac{1}{\surd 2}(\textbf{e}_\rho- i \sigma e^{ik_0\Delta n z} \textbf{e}_\varphi) 
$$
where
$$\Delta n(\rho)=n_\varphi-n_\rho=\frac{\beta (p_{11}-p_{12})}{2\rho}n^3
$$
is the induced birefringence. Equivalently,
\begin{equation}
\textbf{e}_\sigma =\frac{1}{\surd 2}(\textbf{e}_\rho- i \sigma e^{i\gamma \cot \theta} \textbf{e}_\varphi) \label{e}
\end{equation}
where $\cot \theta=z/\rho$ and $\gamma=\frac{1}{2}\beta k_0 (p_{11}-p_{12})n^3$. 

Equations (\ref{soln}) to (\ref{e}) form our expression for the field of a paraxial beam propagating in an optically weakly inhomogeneous and isotropic amorphous medium with screw dislocation. With the beam scalar product defined by \cite{Allen3}
$$
(\textbf{E}_{\sigma lm}|\textbf{E}_{\sigma' l'm'})=\int \int \textbf{E}_{\sigma lm}^\dagger \textbf{E}_{\sigma' l'm'}\; \rho d\rho d\varphi=\delta_{\sigma \sigma'} \delta_{mm'} \int R_{l|m|}^\star R_{l'|m|}\; 2\pi\rho d\rho 
$$
these beams form a complete orthonormal set, in terms of which an arbitrary field distribution can be constructed. The paraxial beam contains an optical dislocation, namely the optical vortex of strength $m$ along its axis, which is a topological object on the wavefront the beam \cite{Berry2}. The $z$-component of the beam's orbital angular momentum (per photon) is given by the expectation value 
$$
\int\int E_{lm}^\star(-i\partial_\varphi) E_{lm}\;\rho d\rho d\varphi=m+\beta k.
$$
The fact that its value has changed from $m$ to $m+\beta k$ owes itself to the torque exerted by the strain field of the dislocation. The effect of the screw dislocation on the orbital angular momentum, therefore, is to change the optical vortex strength, rendering vortex annihilation (for negative values of $m$, i.e., left handed helical modes) or generation of non-integral vortices \cite{Gotte} possible. This is useful in generating fast switchable helical modes for optical information encoding \cite{Marrucci,Marrucci2}. Moreover, the $z$-component of the beam's spin angular momentum is \cite{Berry3}
$$
\int\int -i \textbf{e}_z\cdot\textbf{E}_{\sigma lm}^\star\times\textbf{E}_{\sigma lm}\;\rho d\rho d\varphi=-i \textbf{e}_z\cdot\textbf{e}_\sigma^\star\times\textbf{e}_\sigma=\sigma \cos(\gamma \cot \theta).
$$
As expected \cite{Berry3}, the spin component varies along the paraxial direction due to the induced elasto-optic birefringence. Therefore, a screw dislocated medium exerts torques that change both the spin and orbital parts of the beam's angular momentum. The spin change, being attributed to the elasto-optic effect, is more pronounced for beams that propagate close to the dislocation line. Note that, the $z$-component of the beam's total angular momentum, 
$$
j=\sigma \cos(\gamma \cot \theta)+m+\beta k
$$
reduces to the constant value $\sigma+m$ in the absence of the screw dislocation ($\beta=0$), as it should. 

\section{Paraxial beam dynamics in the dislocated medium}

The adiabatic variation of the wave number $k$, caused by the weak inhomogeneity, has negligible dynamical effect and was, therefore, ignored. However, the adiabatic variation of the beam direction (given by its wave vector $\textbf{k}$) plays a geometric role with nontrivial consequences for the beam dynamics. To incorporate this variation, we treat the coordinate frame used in the previous section as a local frame following the beam direction. As usual, the variation gives rise to a parallel transport law in the momentum space, defined by the Berry connection (gauge potential)
$$
\textbf{A}_{\sigma \sigma'}(\textbf{k})=(\textbf{E}_{\sigma lm}|-i\nabla_\textbf{\scriptsize k}| \textbf{E}_{\sigma' lm}).
$$
Using (\ref{soln}) to (\ref{e}), we obtain
$$
\textbf{A}_{\sigma \sigma'}=[j\delta_{\sigma \sigma'}+i\sigma (\delta_{\sigma \sigma'}-1) \sin(\gamma \cot \theta)]\frac{\cot\theta}{k} \textbf{e}_\varphi
$$
or, in matrix notation,
\begin{equation}
\hat{\textbf{A}}=(m+\beta k+ \hat{{\bf \Sigma}} \cdot\textbf{h})\frac{\cot\theta}{k} \textbf{e}_\varphi \label{m}
\end{equation}
where $\hat{{\bf \Sigma}}=(\hat{\sigma}_1,\hat{\sigma}_2,\hat{\sigma}_3)$ is the Pauli matrix vector and
$$
\textbf{h}(\theta)=(0,\sin(\gamma \cot \theta),\cos(\gamma \cot \theta)).
$$
Equation (\ref{m}) generalizes the result of \cite{Bliokh4} for dislocated (amorphous) media. The first two terms are associated with the parallel transport of the beam's transverse structure \cite{Bliokh4}, while the last describes the parallel transport of the polarization vector along the beam. Note, in particular, the appearance of $\hat{\sigma}_2$ which shows the non-Abelian nature of the polarization transport, to be anticipated in view of the anisotropy \cite{Bliokh5}. The Berry curvature (gauge field strength) associated with this connection is ($\hat{\textbf{A}} \times \hat{\textbf{A}}=0$)
$$
\hat{\textbf{B}}=\nabla_\textbf{\scriptsize k} \times \hat{\textbf{A}}=-(m+\beta k+ \hat{{\bf \Sigma}} \cdot \textbf{D})\frac{\textbf{k}}{k^3} 
$$
where $\textbf{D}(\theta)= d(\textbf{h}\cos\theta)/d(\cos\theta)$. 

In the course of propagation, the paraxial beam evolves according to $\textbf{E}_{\sigma lm}\rightarrow e^{i\hat{\Theta}}\textbf{E}_{\sigma lm}$, where  
\begin{equation}
\hat{\Theta}=\int_C \hat{\textbf{A}} \cdot d\textbf{k}= (m+\beta k)\Theta_0+ \hat{{\bf \Sigma}} \cdot \textbf{I} \label{ph}
\end{equation}
is the geometric Berry phase. Here $C$ is the beam trajectory in momentum space, $\Theta_0=\int_C \cos \theta d\varphi$ is the Berry phase accumulated for $\sigma=1$ and $m=0$ in the absence of dislocation and $\textbf{I}=\int_C \textbf{h} \cos\theta d\varphi=(0,\text{Im} \int_C e^{i\gamma\cot\theta}\cos\theta d\varphi,\text{Re} \int_C e^{i\gamma\cot\theta}\cos\theta d\varphi)$. The first term yields a rotation (through angle -$\Theta_0$) of the beam's transverse structure about the direction of propagation, while the second yields a polarization precession about the direction of $\textbf{I}$. Such a spin precession is characteristic of anisotropic media \cite{Bliokh5}. (In the absence of dislocation, the second term in (\ref{ph}) simply yields the phase factor $e^{i\sigma\Theta_0}$ that leads to the well known Rytov rotation.)

In view of the polarization evolution, the Berry curvature for a given beam is, therefore,
$$
\textbf{B}=(e^{i\hat{\Theta}}\textbf{e}_\sigma)^\dagger\hat{\textbf{B}} (e^{i\hat{\Theta}}\textbf{e}_\sigma)=
\textbf{e}_\sigma^\dagger\hat{\textbf{B}}'\textbf{e}_\sigma
$$
where
\begin{eqnarray}
\hat{\textbf{B}}'=e^{-i\hat{\Theta}}\hat{\textbf{B}} e^{i\hat{\Theta}}=-(m+\beta k+\hat{{\bf \Sigma}}' \cdot\textbf{D})\frac{\textbf{k}}{k^3}\nonumber \\ 
\hat{{\bf \Sigma}}'=\exp(-i\hat{{\bf \Sigma}} \cdot \textbf{I}) \hat{{\bf \Sigma}}\exp(i\hat{{\bf \Sigma}} \cdot \textbf{I}). \ \ \ \ \ \nonumber
\end{eqnarray}
Using the Baker-Hausdorff formula,
$$
e^{-iG}Ae^{iG}=A-i[G,A]+\frac{(-i)^2}{2!}[G,[G,A]]+\frac{(-i)^3}{3!}[G,[G,[G,A]]]+\ldots
$$
and $[\sigma_i ,\sigma_j]=2i\epsilon_{ijk}\sigma_k$, after some calculations we find
$$
\hat{{\bf \Sigma}}'=\hat{{\bf \Sigma}}\cos 2I+\frac{\textbf{I}(\hat{{\bf \Sigma}}\cdot \textbf{I})}{I^2}(1-\cos 2I)+\frac{\hat{{\bf \Sigma}}\times \textbf{I}}{I} \sin 2I
$$
where $I=|\textbf{I}|=|\int_C e^{i\gamma\cot\theta}\cos\theta d\varphi|$.
Hence
$$
\hat{\textbf{B}}'=-(m+\beta k+\hat{{\bf \Sigma}} \cdot\textbf{D}')\frac{\textbf{k}}{k^3}
$$
where
$$
\textbf{D}'=\textbf{D}\cos 2I+\frac{\textbf{I}(\textbf{D}\cdot \textbf{I})}{I^2}(1-\cos 2I)-\frac{\textbf{D}\times \textbf{I}}{I} \sin 2I.
$$
Therefore,
$$
\textbf{B}=-(m+\beta k+\sigma D'_3)\frac{\textbf{k}}{k^3}
$$
with
$$
D_3'=D_3\cos 2I+\frac{I_3(\textbf{D}\cdot \textbf{I})}{I^2}(1-\cos 2I)
$$
which reduces to the result of \cite{Bliokh4} in the absence of dislocation, namely, the field of a magnetic monopole of charge $m+\sigma$ situated at the origin of the momentum space. 

The equations of motion of the beam in the presence of momentum space Berry curvature have been derived repeatedly for various particle beams (photons \cite{Bliokh,Bliokh1,Bliokh2,Zeldovich2,Onoda,Onoda2,Sawada}, phonons \cite{Bliokh6,Torabi,Mehrafarin} and electrons \cite{Chang,Sundaram,Culcer,Berard}). We have
\begin{eqnarray}
\dot{\textbf{k}}=k\nabla \ln n \ \ \nonumber\\
\dot{\textbf{r}}=\frac{\textbf{k}}{k}+\textbf{B}\times \dot{\textbf{k}}\nonumber
\end{eqnarray}
where dot denotes derivative with respect to the beam length. These differ from the standard ray equations of the geometrical optics by the term involving the Berry curvature, which yields the beam displacement
\begin{equation}
\delta \textbf{r}=-\int_C (m+\beta k+\sigma D'_3)\frac{\textbf{k}\times d\textbf{k}}{k^3}. \label{def}
\end{equation}
The displacement, which results in the splitting of beams with different polarizations and/or orbital angular momentums, is orthogonal to the beam direction and produces a current across the direction of propagation. This is the optical Hall effect in the dislocated medium and generalizes the result of \cite{Bliokh4}. Because the beam's singular vortex core can be observed with great accuracy, measurement of the displacements is very reliable. Measuring $\delta\textbf{r}$ for opposite values of $m$ and $\sigma$ provides an indirect method for determining the Burgers vector as well as the elasto-optic coefficients (or rather their difference $p_{11}-p_{12}$) of the medium.

\section{Example: circular waveguide}

We conclude by describing a simple situation that demonstrates the application of the result (\ref{def}) in determining the magnitude of the Burgers vector. 

\begin{figure}
\includegraphics{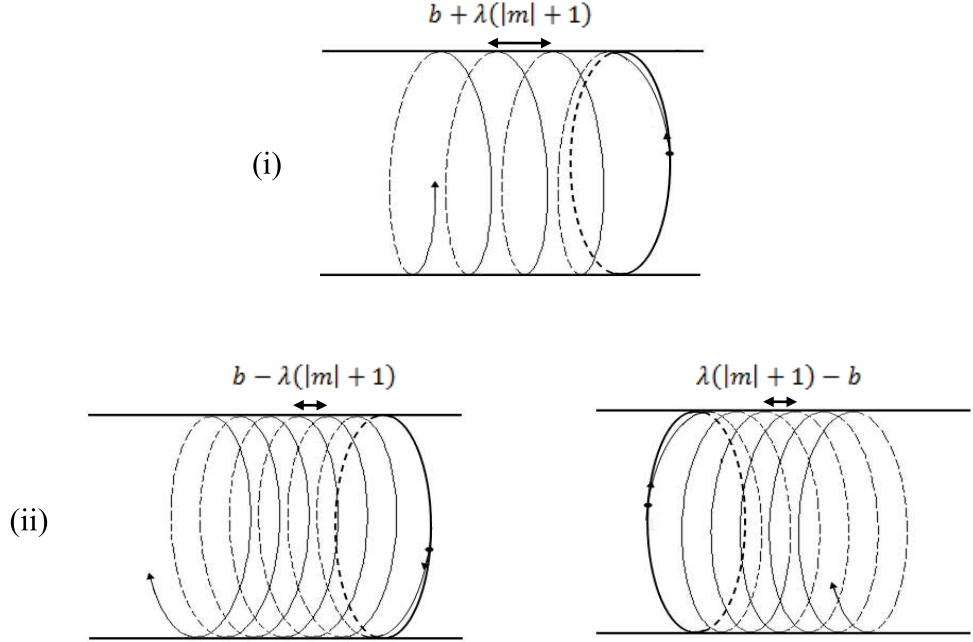}	
\caption{
Propagation of oppositely polarized beams inside a screw dislocated circular waveguide. (i) $\sigma=1, m=|m|$, (ii) $\sigma=-1, m=-|m|$ with $\lambda (|m|+1)<b$ (left) and $\lambda (|m|+1)>b$ (right).
\label{fig1}}
\end{figure}

Silicon, being transparent in the near infrared, has attracted extensive attention in optoelectronic devices. Waveguides in silicon can serve as optical interconnects in silicon integrated circuits, or distribute optical clock signals in a microprocessor. Fabrication of such waveguides requires a core with a higher refractive index than that of crystalline silicon.  Amorphous silicon, with a higher refractive index in the near infrared, thus provides an interesting candidate for the
core material \cite{Dood}. Let us, therefore, consider an amorphous silicon cylindrical waveguide with screw dislocation along its axis $z$. For a beam propagating inside the waveguide along its circular cross section, the trajectory $C$ in (\ref{def}) is a circle of radius $k=2\pi/\lambda$ in the $xy$-plane ($\theta=\pi/2$). Thus, $I=0$ and $D'_3=D_3=1$ so that (\ref{def}) yields, for every one revolution of the beam,
$$
\delta \textbf{r}=-[b+\lambda (m+\sigma)] \textbf{e}_z
$$
where, of course, $\textbf{e}_z$ is the unit vector in the $z$-direction and $b=2\pi \beta$ is the magnitude of the Burgers vector. Therefore, for two beams of opposite polarization and orbital angular momentum, we have a displacement (per unit revolution) in the negative z-direction given by 
$$
\delta z^\pm=b\pm\lambda (|m|+1). 
$$
While $\delta z^+>0$, $\delta z^-$ can take any value (including zero) by suitably tuning $\lambda$ and $|m|$. Thus one can have the oppositely polarized beams emerging from the same end of the waveguide, with one lagging behind the other, or from opposite ends (see FIG.~\ref{fig1}). Keeping the emerging direction of the right-handed beam as the reference direction, the emerging direction of the left handed beam changes at the threshold value $\lambda (|m|+1)=b$. With $\lambda$ in the near infrared, this provides an experimental method of determining Burgers vector magnitudes of the order of micrometers.

\end{document}